# Lifting of *xz/yz* orbital degeneracy at the structural transition in detwinned FeSe

T. Shimojima[1,*], Y. Suzuki[1], T. Sonobe[1], A. Nakamura[1], M. Sakano[1], J. Omachi[2],
K. Yoshioka[3], M. Kuwata-Gonokami[2,3], K. Ono[4], H. Kumigashira[4], A. E. Böhmer[5],
F. Hardy[5], T. Wolf[5], C. Meingast[5], H. v. Löhneysen[5,6], H. Ikeda[7], K. Ishizaka[1]

[1]*Quantum-Phase Electronics Center(QPEC) and Department of Applied Physics, University of Tokyo, Bunkyo, Tokyo 113-8656, Japan.*
[2]*Photon Science Center, The University of Tokyo, 7-3-1 Hongo, Bunkyo-ku, Tokyo 113-8656, Japan*
[3]*Department of Physics, The University of Tokyo, 7-3-1 Hongo, Bunkyo-ku, Tokyo 113-0033, Japan*
[4]*KEK, Photon Factory, Tsukuba, Ibaraki 305-0801, Japan.*
[5]*Institut für Festkörperphysik, Karlsruhe Institute of Technology, 76021 Karlsruhe, Germany*
[6]*Physikalisches Institut, Karlsruhe Institute of Technology, 76128 Karlsruhe, Germany*
[7]*Department of Physics, Kyoto University, Kyoto 606-8502, Japan.*

We study superconducting FeSe ($T_c$ = 9 K) exhibiting the tetragonal-orthorhombic structural transition ($T_s$ ~ 90 K) without any antiferromagnetic ordering, by utilizing angle-resolved photoemission spectroscopy. In the detwinned orthorhombic state, the energy position of the $d_{yz}$ orbital band at the Brillouin zone corner is 50 meV higher than that of $d_{xz}$, indicating the orbital order similar to NaFeAs and BaFe$_2$As$_2$ families. Evidence of orbital order also appears in the hole bands at the Brillouin zone center. Precisely measured temperature dependence using strain-free samples shows that the onset of the orbital ordering ($T_o$) occurs very close to $T_s$, thus suggesting that the electronic nematicity above $T_s$ is considerably weaker in FeSe compared to BaFe$_2$As$_2$ family.

The parent compounds of iron-based superconductors [1] mostly exhibit complex phase transitions characterized by the antiferromagnetic (AFM) transition at $T_N$ occurring simultaneously with or just below the tetragonal-orthorhombic structural transition at $T_s$ [2]. These phase transitions are similarly suppressed toward the appearance of the superconducting (SC) phase by partial chemical substitution or applying pressure [3], indicating a strong relation between magnetism, superconductivity, and lattice deformation. Recently, systematic x-ray diffraction and torque magnetometry measurements on P-doped BaFe$_2$As$_2$ (Ba122) further revealed an in-plane anisotropy developing at $T^*$ above $T_s$ and $T_N$ [4], i.e., an electronic nematicity. The nematic susceptibility over the whole SC regions in Co-doped and K-doped Ba122 extracted from measurements of the elastic shear modulus [5], as well as high-resolution thermal expansion data [6], on the other hand do not show any anomalies suggestive of a four-fold symmetry reduction above $T_s$ [5].

In association with the abovementioned complex phase diagram, theoretical studies raised the possibility of orbital ordering [7-9] characterized by the nonequivalent electronic occupation in $d_{xz}$ ($xz$) and $d_{yz}$ ($yz$) orbitals, which may couple the AFM and lattice deformation. Polarization-dependent laser-based angle-resolved photoemission spectroscopy (ARPES) indeed reported a nearly orbital-polarized (either $xz$ or $yz$) Fermi surfaces (FSs) in the AFM state of non-doped Ba122, suggestive of an orbital-ordered state [10]. Later, ARPES study on lightly Co-doped[11] and P-doped Ba122[12] reported the nonequivalent shifts of the $xz/yz$ orbitals appearing above $T_s$, thus suggesting the possible complex interplay between the electronic nematicity and the orbital anisotropy. Regarding the theoretical origin of the electronic nematicity, it is based on either magnetism [13-17] or orbital degrees of freedom [7-9,18]. In the former, $Z_2$ spin-nematic ordering induces the orthorhombic lattice distortion and $xz/yz$ orbital order. On the other hand, the latter scenario defines the order parameter of the ferro-orbital ordering as $|N_{xz} - N_{yz}|$, where $N_{xz}$ and $N_{yz}$ represent the number of occupied Fe $3d_{xz}$ and $d_{yz}$ states below the Fermi level ($E_F$). $|N_{xz} - N_{yz}|$ naturally reduces the symmetry from four-fold to two-fold, leading to the orthorhombic transition and the FS nesting-driven AFM state [18].

The electronic structures of the respective phases thus have been investigated both theoretically and experimentally. Nevertheless, it is difficult to separately distinguish and evaluate the effects of AFM and orbital ordering, since in most cases they are closely coupled to each other in a narrow temperature range. Recently, NaFeAs (Na111) possessing fairly well separated transition temperatures of $T_s$ = 53 K and $T_N$ = 40 K, has been investigated by ARPES [19,20]. In Na111, the orbital order was found to develop just at or slightly above $T_s$, which seems to trigger the stripe AFM order at lower $T_N$, due to the evolution of the two-fold anisotropic FSs [19,20]. Thus, an intimate relation between orbital and AFM ordering was explained based on the orbital-driven scenario [19].

On the other hand, FeSe is a peculiar system among iron-based superconductors. It has the simplest crystal structure and undergoes a tetragonal-orthorhombic structural transition at around 90 K without any AFM ordering [21]. The absence of the AFM transition in FeSe poses some questions on the existence of the orbital ordering and its relation to the origin of the structural transition. Precise ARPES measurements on FeSe will offer valuable insight for the interplay between the electronic and the lattice degrees of freedom, which seems to strongly depend on the material class in the

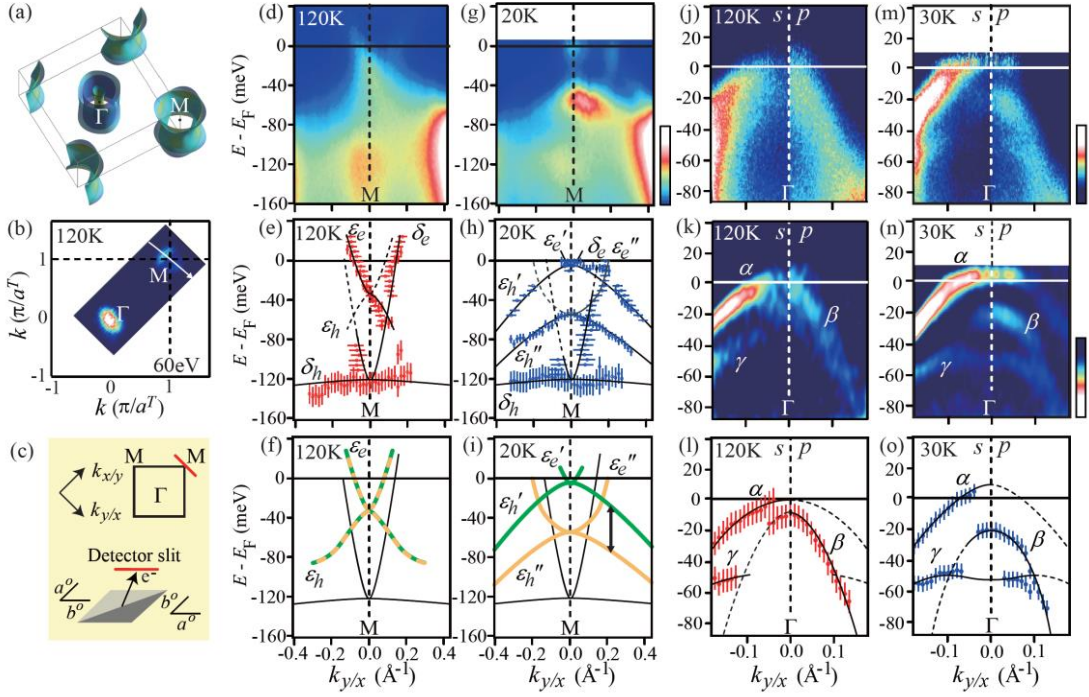

Figure 1 (a) Calculated FSs in the tetragonal phase of FeSe. (b) FSs in the tetragonal phase detected by synchrotron-radiation ARPES with the circular-polarized photons of $h\nu = 60$ eV. White arrow indicates the momentum cut in (d-i). $a^T$ represents the in-plane lattice parameter of the tetragonal structure. (c) Schematics of the twinned domains in the orthorhombic phase, where M along $k_x$ and M along $k_y$ are overlapped below $T_s$. (d) ARPES image ($h\nu = 60$ eV) at the BZ corner for 120 K ($T_s < T$) divided by the Fermi-Dirac (FD) function including the energy resolution, (e) MDC and EDC peak plots with the band assignment, and (f) the schematically drawn band dispersions. Black curves in (e) indicate the band dispersions estimated from the peak positions, whereas the broken black curves are those symmetrized with respect to the M point. (g-i) Similar to (d-f) recorded at 20 K ($T < T_s$). Orange and green curves in (f,i) represent the $xz$ and $yz$ orbitals, respectively. Black arrow indicates the relative energy between $xz$ and $yz$ orbitals. (j) ARPES image ($h\nu = 5.9$ eV) at the BZ center for 120 K ($T_s < T$) divided by FD function including the energy resolution, (k) its second $E$-derivative, and (l) the band dispersions with peak plots. Left (right) side in (j) and (k) show the ARPES data taken by $s(p)$-polarization. (m-o) Similar to (j-l) recorded at 30 K ($T < T_s$).

iron-based superconductors.

    In this work, the electronic structure of FeSe in a wide range of temperature ($T$) is investigated by using laser, He-discharge lamp, and synchrotron-radiation ARPES. Nonequivalent energy shifts of $xz/yz$ orbital bands are found to arise below $T_s$. The result from detwinned sample clearly shows that the $yz$ orbital is located at 50 meV higher energy compared to $xz$, at the Brillouin zone (BZ) corners. This energy difference of the $xz/yz$ orbitals at the BZ corners is more than five times larger than expected by DFT calculation considering only the orthorhombic lattice distortion. It thus suggests an electronic origin of the nonequivalent $xz/yz$ orbitals, namely orbital order. The good correspondence between $T_s$ and $T_o$ indicates that the structural transition in FeSe is simultaneously driven by the orbital ordering, without any strong indication of electronic nematicity above $T_s$.

    Single crystals of FeSe were synthesized from Fe and Se powders, which were mixed in an atomic ratio 1.1:1 and sealed in an evacuated $SiO_2$ ampoule together with a eutectic mixture of KCl and $AlCl_3$, as described in Ref. 21. The structural transition and SC transition temperatures of the single crystals used in the present work were estimated to be $T_s \sim 90$ K and $T_c = 9.8$ K from the resistivity. ARPES measurements at the BZ center were performed using VG-Scienta R4000WAL electron analyzer and the 4th-harmonic generation of Ti-Sapphire laser radiation ($h\nu = 5.9$ eV) [22] at University of Tokyo [23]. $s$- and $p$-polarizations were used. The energy resolution was set to 3 meV. ARPES measurements at the BZ corner were performed using VG-Scienta SES2002 and synchrotron radiation of 60 eV at BL 28A of Photon Factory. Circular-polarized light was used with the total energy resolution of 20 meV. For the ARPES measurements on the detwinned single crystals, we used VG-Scienta R4000WAL and a Helium discharge lamp of $h\nu = 21.2$ eV at University of Tokyo. The energy resolution was set to 10 meV. In order to detwin the single crystals, we applied the tensile strain along one of the tetragonal ($\pi,\pi$) directions. Below $T_s$, the tensile strain brings the orthorhombic $a^o$ axis ($a^o > b^o$) along its direction. The crystals were cleaved *in situ* at $T = 10$ K in an ultrahigh vacuum of $7 \times 10^{-11}$ Torr. Electronic band structure calculations were performed based on density functional theory (DFT) using the relativistic full potential (linearized) augmented plane-wave (FLAPW) +

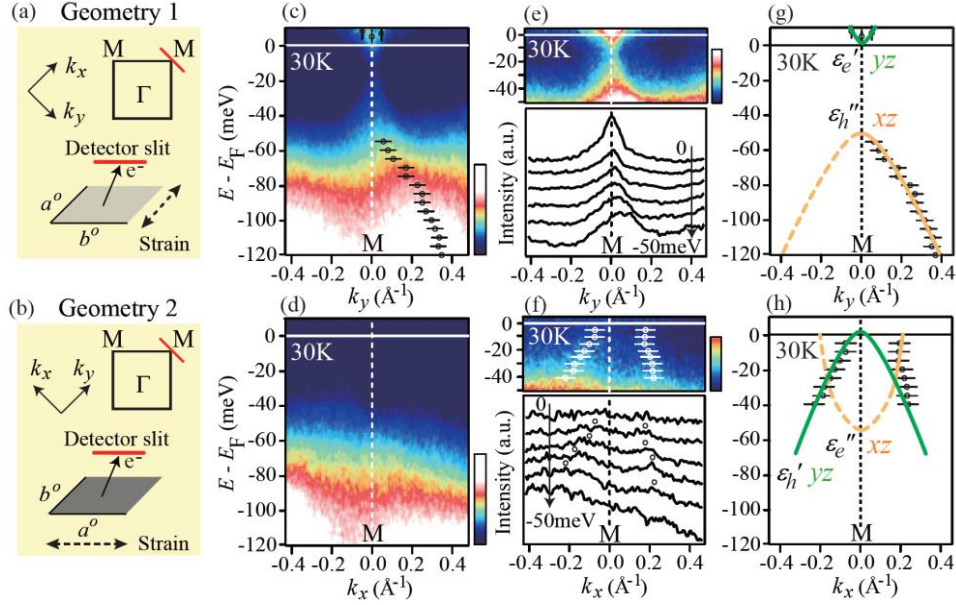

Figure 2 (a,b) Two experimental geometries for ARPES ($h\nu = 21.2$ eV) on detwinned samples. The direction of the tensile strain is shown by the broken arrows, with respect to the detector slit. Band dispersions at the M point along $k_y$ ($k_x$) are detected in Geometry 1 (2). $x$ and $y$ correspond to the orientations of $a^o$ and $b^o$, respectively. (c,d) ARPES images divided by the FD function including the energy resolution, taken at M along $k_y$ and M along $k_x$, respectively. (e,f) Upper panels show the high-contrast ARPES image near $E_F$ at M along $k_y$ and M along $k_x$, respectively. Lower panels show the MDCs from $E_F$ to -50 meV. (g,h) MDC and EDC peak plots and the schematic band dispersions at M along $k_y$ and M along $k_x$, respectively. Orange and green curves represent $xz$ and $yz$ orbital band, respectively.

local orbitals method as implemented in the WIEN2k package [24].

The band calculation of FeSe in the tetragonal phase shows quasi-two dimensional FSs as depicted in Fig. 1(a). The mapping of ARPES intensity at $E_F \pm 5$ meV for strain-free FeSe [Fig. 1(b)], measured at 120 K and $h\nu = 60$ eV, shows FSs similar to the calculation; the hole and electron FSs are located at the tetragonal BZ center ($\Gamma$) and corners (M), respectively. Band dispersions in both tetragonal [120 K, Fig. 1(d)] and orthorhombic [20 K, Fig. 1(g)] phases were taken along the momentum cut indicated by the white arrow in Fig. 1(b). Note that since the crystal at $T < T_s$ is composed of twinned domains of the orthorhombic structure, M along $k_x$ and M along $k_y$ become overlapped in the ARPES image as schematically shown in Fig. 1(c).

Figures 1(d) and (e) show the ARPES image at 120 K and corresponding band dispersion obtained by plotting the peaks of momentum distribution curves (MDCs, with horizontal error bars) and energy distribution curves (EDCs, with vertical error bars). One can identify two electron bands ($\delta_e$, $\varepsilon_e$) and two hole bands ($\delta_h$, $\varepsilon_h$) around the M point. According to the band calculations, $\delta$ and $\varepsilon$ bands are mainly composed of $xy$ and $xz/yz$ orbital character, respectively. The observed inner and outer electron FS pockets around the M point are common to the Ba122 family. With decreasing $T$, the band dispersions exhibit drastic changes in both number and energy position as shown in Fig. 1(g). The most prominent features appearing at $T = 20$ K are the shallow electron-like band with its bottom just below $E_F$ and the hole-like band showing its maximum at around 50 meV below $E_F$ (i.e., $E - E_F = -50$ meV), at the M point. Some additional band dispersions are also recognized at 0.2 – 0.4 Å$^{-1}$ away from the M point. By plotting the peaks of EDCs and MDCs, and by assuming a symmetric dispersion with respect to the M point, we can identify three electron bands ($\delta_e$, $\varepsilon_e'$, $\varepsilon_e''$) and three hole bands ($\delta_h$, $\varepsilon_h'$, $\varepsilon_h''$) at 20 K, as depicted in Fig. 1(h). It is worth noting that the $\delta_h$ and $\delta_e$ bands composed of $xy$ orbitals are almost $T$-independent.

The increase in the number of $\varepsilon_e$ and $\varepsilon_h$ bands is consistently explained by considering both the lifting of degeneracy in $xz/yz$ bands (i.e., orbital order) and the formation of the twinned domains below $T_s$. As already discussed in the literature [11,12], orbital ordering in twinned crystals doubles the number of observed $xz/yz$ bands around the BZ corners, due to the overlap of nonequivalent M along $k_x$ and M along $k_y$. We can thus derive the $xz/yz$ orbital order appearing as the splitting of $\varepsilon$ hole and electron bands, as schematically shown in Fig. 1(i). By comparing the obtained band dispersions at 120 K [Fig. 1(e)] and 20 K [Fig. 1(h)], the $\varepsilon_e'$ band shifts upward by 25 meV whereas the $\varepsilon_e''$ band moves downward by 25 meV, with respect to the original $\varepsilon_e$ band in the tetragonal phase.

The impact of the orbital order also shows up at the BZ center. Around the $\Gamma$ point in the tetragonal phase (120 K), as shown in Fig. 1(j), there are three hole bands ($\alpha$, $\beta$, $\gamma$) observed with different gradients. The

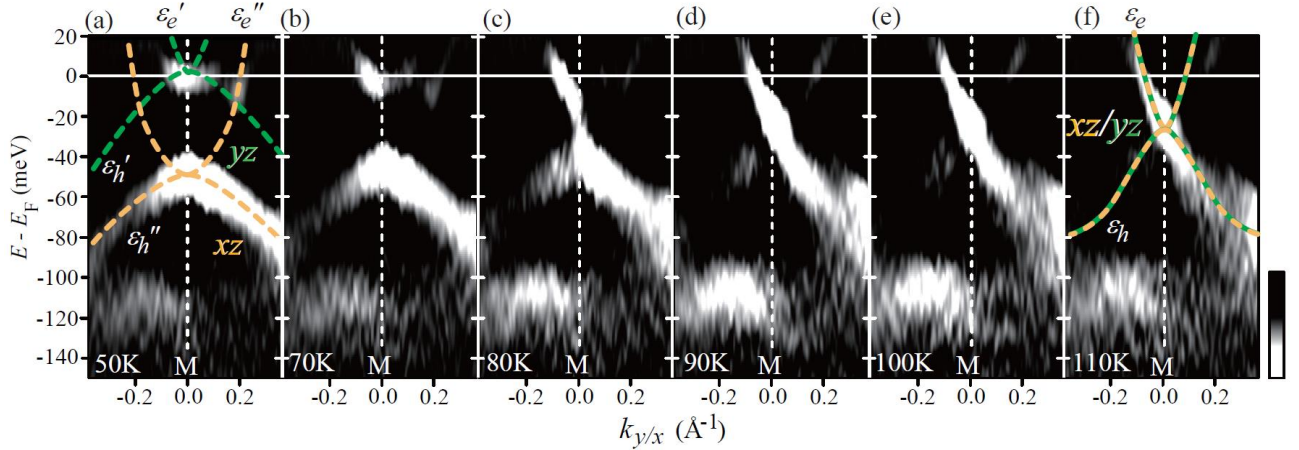

Figure 3 (a)-(f) $T$ dependence of the second $E$-derivative of ARPES images taken at 50, 70, 80, 90, 100, and 110 K at twinned M point ($h\nu = 60$ eV). Orange and green broken curves represent $xz$ and $yz$ orbital band, respectively.

second $E$-derivative of the ARPES image [Fig. 1(k)] emphasizes that $\alpha$ and $\gamma$ bands ($\beta$ band) are clearly detected by $s$ ($p$)-polarization configuration. The band dispersions obtained from the EDC peaks in Fig.1 (l) show that the $\alpha$ and $\beta$ bands are nearly degenerate at the BZ center, whereas the $\gamma$ band is located at around $E - E_F = -50$ meV. At 30 K, the multiple structure of the bands at the BZ center becomes clearer [Fig. 1(m,n)]. The energy positions of the $\alpha$, $\beta$ and $\gamma$ hole bands in the orthorhombic phase, as shown in Fig. 1(o), are consistent with previous ARPES on FeSe [25,26]. Ref.25 reported that the $\alpha$ and $\beta$ hole bands predominantly include $xz/yz$ orbital character. The opposite polarization dependence of $\alpha$ and $\beta$ bands [Fig.1 (m,n)] is also consistent with this interpretation. By comparing the second $E$-derivative images in Fig.1 (k) and (n), a clear $T$ dependence for these $\alpha$ and $\beta$ bands is recognized. As schematically shown in Fig.1 (o), the separation between $\alpha$ and $\beta$ bands at the $\Gamma$ point at 30 K is estimated to be ~30 meV, in contrast to the value found at 120 K. The energy scale of the electronic modification is nearly comparable to that for the $xz/yz$ orbital order at the BZ corners.

In order to conclusively assign the $xz$ and $yz$ orbital component at the BZ corners, we detwinned the single crystals by applying tensile strain along one of the tetragonal ($\pi,\pi$) direction. The tensile strain brings the orthorhombic $a^o$ axis along its direction, at $T < T_s$. As illustrated in Figs. 2(a) and (b), M along $k_y$ and M along $k_x$ points are thus separately recorded by using Geometries 1 and 2. The obtained ARPES images at M along $k_y$ and M along $k_x$ are shown in Figs. 2(c) and (d), respectively. In Fig. 2(c), the electron-like $\varepsilon_e'$ band touching $E_F$ and the hole-like $\varepsilon_h''$ band at $E - E_F = -50$ meV can be identified from the peak positions of the EDCs and MDCs, while they are absent in Fig.2 (d). In the enhanced color scale as shown in Figs. 2(e) and (f), on the other hand, a hole-like $\varepsilon_h'$ band approaching $E_F$ becomes visible only in Geometry 2. These observations of band dispersions around M along $k_y$ and M along $k_x$ are summarized in Figs. 2 (g) and (h), by plotting the EDC and MDC peaks. According to the band calculations on FeSe, the $yz$ ($xz$) orbital forms a saddle point at the BZ corners, with the electron-like band along $\Gamma$-M $||k_y$ ($\Gamma$-M $||k_x$) and the hole-like band along $\Gamma$-M $//k_x$ ($\Gamma$-M $//k_y$) direction. The present ARPES data on detwinned single crystals indicates an upward shift of the $yz$ band and a downward shift of the $xz$ orbital band at low $T$. We mention that the direction of the energy shift in the $xz/yz$ orbital bands of FeSe is also common to the Ba122 and Na111 systems.

Now we move on to the detailed $T$ dependence of the band dispersions at the BZ corner across $T_s$. Here we used the strain-free twinned single crystals in order to extract the intrinsic $T$ dependence, without the influence of uniaxial pressure. Figures 3(a-f) show the second $E$-derivative of the ARPES image at the M point. In the image at 50 K as shown in Fig. 3(a), the $\varepsilon_e'$, $\varepsilon_e''$ and $\varepsilon_h''$ bands are clearly emphasized. By tracking the $\varepsilon_e'$ and $\varepsilon_h''$ bands, respectively representing the nonequivalent $yz$ and $xz$ orbitals, a drastic change appears at 90 K [Fig. 3(d)]. Above 90 K, the overall band dispersions are almost $T$-independent and the $xz/yz$ orbital bands become degenerate.

In order to precisely estimate the onset $T$ of the $xz/yz$ orbital ordering, the EDCs at the M point are shown for 20 - 140 K in Fig. 4 (a). The peak positions of the EDCs can be determined by the bottom positions of the second-derivative curves (not shown). At 20 K, the peaks of the EDC are located at 3, 55, and 115 meV, corresponding to $\varepsilon_e'$($yz$), $\varepsilon_h''$($xz$) and $\delta_h$ ($xy$) bands, respectively. With increasing $T$, the two peaks of $\varepsilon_e'$ and $\varepsilon_h''$ get closer and merge into a single peak at ~90 K, in contrast to the $\delta_h$ band ($xy$), which remains nearly $T$-independent. The peak energies of EDC for the $\varepsilon_e'$ and $\varepsilon_h''$ bands, $E_{yz}$ and $E_{xz}$, are plotted in Fig. 4(d). This clearly indicates that the energetically nonequivalent $xz$ and $yz$ orbitals become degenerate at 90 K.

The order parameter of the orbital order, as we here define by $\Phi_o(T) = E_{yz} - E_{xz}$, shows the value of $\Phi_o \sim$ 50 meV at the lowest $T$ and $T_o \sim 90$ K [Fig. 4(e)].

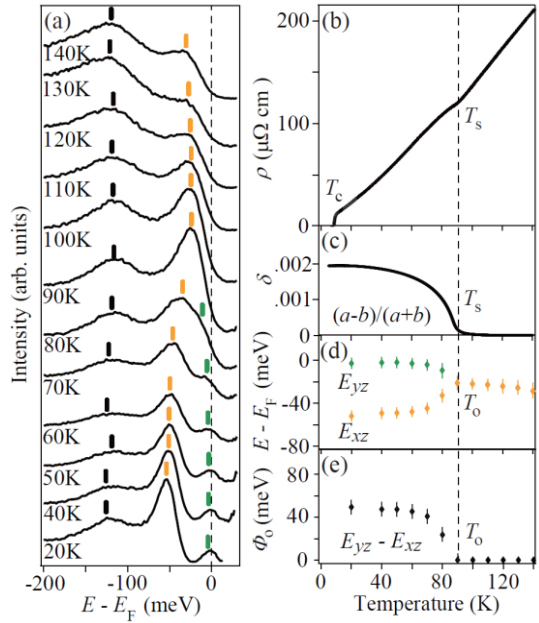

Figure 4(a) $T$ dependence of the EDCs at the M point divided by the FD functions including the energy resolution. Peak energies of $\varepsilon_e'$ ($yz$), $\varepsilon_h''$ ($xz$), and $\delta_h$ ($xy$) bands at the M point are denoted by green, orange, and black markers. (b) $T$ dependence of the in-plane resistivity $\rho$, (c) orthorhombicity $\delta(T) = (a - b)/(a + b)$ [21], (d) energy of $xz$ and $yz$ orbital bands $E_{xz}$ and $E_{yz}$, and (e) the orbital order parameter $\Phi_o(T) = E_{yz} - E_{xz}$.

$\Phi_o \sim 50$ meV is significantly larger than expected if driven only by the orthorhombic lattice distortion (estimated to be < 10 meV by DFT calculation for FeSe). This indicates that the value $\Phi_o \sim 50$ meV is evidence of an electronically driven orbital order. On the other hand, $T_s$ of the FeSe single crystal was estimated to be ~90 K by electrical resistivity [Fig. 4(b)] and thermal expansion measurements [Fig. 4(c)] [21]. The close correspondence between $T_o$ and $T_s$ may suggest that the structural transition itself is triggered by orbital order. At the same time, it also indicates that the electronic nematicity above $T_s$ is severely weaker in FeSe compared to Ba122 family.

Here we find that 122, 111 and 11 families show orbital ordering of similar behavior, i.e., $E_{yz} > E_{xz}$ at the BZ corners, on cooling. It is also worth mentioning that the orbital order always appears at the highest $T$ of all phase transitions. These facts suggest that the orbital ordering may be a ubiquitous feature in iron-based superconductors that drives the two-fold symmetric properties in the FeAs or FeSe layers. On the other hand, the relation among $T_s$, $T_N$ and $T_o$ is considerably material-dependent. How the orbital order triggers the structural and magnetic transitions, in some case appearing as the electronic nematicity at higher $T$, seems to depend strongly on the specific material.

Finally we discuss the possible relation between orbital ordering and superconductivity. The present ARPES result, clearly showing the lifting of $xz/yz$ orbital degeneracy, suggests that the shapes of the FSs at the BZ center and corners are modified to a two-fold symmetry. Such a distortion of the FSs below an orbital ordering transition has been also discussed for Na111, as a possible origin of the stripe-type AFM ordering [19]. Theoretical studies including the momentum-independent nonequivalency of $xz/yz$ orbitals with a difference of 40 meV, reported the enhancement of the magnetic susceptibility only along the (0,0) - ($\pi$,0) direction, as compared to the four-fold state [18]. This scenario may explain the enhancement of the spin fluctuations in FeSe, being observed below ~100 K in the orbitally ordered state [27]. Note that the inter-orbital fluctuations can also simultaneously arise from FS nesting between different orbital bands due to the vertex corrections [28]. These spin and/or orbital fluctuations may be playing important roles for superconductivity. Considering its fairly ubiquitous occurrence across the families, this orbital order can be regarded as one of the prerequisite conditions for superconductivity in iron-based superconductors. Further investigation of the FS morphology of FeSe and its nesting condition at low $T$ in the detwinned crystals will be required.

In conclusion, we investigated in detail the electronic structure of superconducting FeSe in a wide range of $T$, across the structural phase transition. By using a detwinned single crystal, the saddle point of the band at the BZ corners consisting of $yz$ orbital was found to be located at 50 meV higher than that of $xz$ orbital, indicating orbital order. The effect of $xz/yz$ orbital order also appeared in the hole bands at the BZ center, which should be taken into account when discussing the FS morphology. From the detailed $T$ dependence, the orbital order was found to show up at $T$ very close to $T_s$, thus implying the orbital origin of the orthorhombic transition in FeSe. By comparing with other families (Ba122 and Na111), the ubiquitous nature of orbital order in iron-based superconductors is suggested, whereas the electronic nematicity above $T_s$ might be considerably weaker for FeSe.

**Acknowledgement,**
We acknowledge Y. Matsuda, T. Shibauchi, S, Kasahara, H. Kontani, S. Onari, K. Okazaki and S. Shin for valuable discussions. Synchrotron-radiation ARPES experiments were carried out at BL-28A at Photon Factory (No. 2012G751). This research was supported by "Toray Science Foundation", "Precursory Research for Embryonic Science and Technology (PRESTO), Japan Science and Technology Agency", "the Photon Frontier Network Program of the MEXT, Japan" and "Research Hub for Advanced Nano Characterization, The University of Tokyo, supported by MEXT, Japan".

*Corresponding author: shimojima@ap.t.u-tokyo.ac.jp